\documentclass[amsmath,amssymb,prl,aps,twocolumn,superscriptaddress]{revtex4-2}
\usepackage{amsmath,amsfonts,amssymb,amsthm,graphics,graphicx,epsfig,bbm}
\usepackage[colorlinks=true,citecolor=blue,linkcolor=blue,urlcolor=blue]{hyperref}
\usepackage[usenames]{color}
\usepackage{graphicx}
\usepackage{subfigure}
\usepackage{amsmath}
\usepackage{epsfig}
\usepackage{dcolumn}
\usepackage{bm}
\usepackage{color}
\usepackage{times}
\usepackage{epstopdf}
\usepackage{amssymb}
\usepackage{amstext}
\usepackage{latexsym}
\usepackage{hyperref}
\usepackage{amsfonts}
\usepackage{psfrag}
\usepackage{soul,xcolor}
\usepackage[normalem]{ulem}
\usepackage{dsfont}
\usepackage{adjustbox}
\usepackage{txfonts}
\usepackage{physics}
\usepackage{float}
\usepackage[english]{babel}
\usepackage[autostyle]{csquotes}
\usepackage{outlines}
\usepackage{amsmath,amsfonts,amssymb,amsthm,graphics,graphicx,epsfig,bbm}
\usepackage[colorlinks=true,citecolor=blue,linkcolor=blue,urlcolor=blue]{hyperref}
\usepackage[usenames]{color}
\usepackage{graphicx}
\usepackage{subfigure}
\usepackage{amsmath}
\usepackage{epsfig}
\usepackage{dcolumn}
\usepackage{bm}
\usepackage{color}
\usepackage{times}
\usepackage{epstopdf}
\usepackage{amssymb}
\usepackage{amstext}
\usepackage{latexsym}
\usepackage{hyperref}
\usepackage{amsfonts}
\usepackage{psfrag}
\usepackage{soul,xcolor}
\usepackage[normalem]{ulem}
\usepackage{dsfont}
\usepackage{adjustbox}
\usepackage{txfonts}
\usepackage{physics}
\usepackage{float}
\usepackage[english]{babel}
\usepackage[autostyle]{csquotes}
\usepackage{outlines}

\newcommand{\inv}[1]{{#1}^{-1}}
\newcommand{\conj}[1]{{#1}^{\dagger}}

\newcommand{\err}{\epsilon}
\newcommand{\I}{\mathds{1}}

\newcommand{\Dc}{\mathcal{D}_{\oplus}}

\newcommand{\C}{\mathcal{C}}

\newcommand{\R}[2]{R^{(#1)}_{#2}}
\newcommand{\K}[2]{K^{(#1)}_{#2}}
\newcommand{\eig}[1]{\lambda_{#1}}

\renewcommand{\L}{\mathcal{L}}
\renewcommand{\O}{\mathcal{O}}

\newcommand{\im}{\iota}

\newcommand{\Ist}{$1^{st}$}
\newcommand{\paper}{letter}

\newcommand{\Map}{\Phi}
\newcommand{\MapI}{\Phi^{(\err)}}
\newcommand{\MapN}{\Phi^{(\err)}_N}

\renewcommand{\l}{\ell}

\renewcommand{\section}[1]{\emph{#1}---}

\begin{document}

\title{Kraus map closed-form solution for general master equation dynamics}
\author{Shahrukh Chishti}
\affiliation{Institute for Theoretical Physics, University of Cologne, D-50937 Cologne, Germany}
\affiliation{Forschungszentrum J\"ulich GmbH, Peter Gr\"unberg Institute, Quantum Control (PGI-8), 52425 J\"ulich, Germany}

\author{F. A. C\'ardenas-L\'opez}
\affiliation{Forschungszentrum J\"ulich GmbH, Peter Gr\"unberg Institute, Quantum Control (PGI-8), 52425 J\"ulich, Germany}

\author{Felix Motzoi}
\affiliation{Institute for Theoretical Physics, University of Cologne, D-50937 Cologne, Germany}
\affiliation{Forschungszentrum J\"ulich GmbH, Peter Gr\"unberg Institute, Quantum Control (PGI-8), 52425 J\"ulich, Germany}

\begin{abstract}
The Kraus representation of quantum channels allows for a precise emulation of the complex dynamics that take place on quantum processors, whether for benchmarking algorithms, predicting the performance of error correction and mitigation, or in the myriad other uses of compiled digital sequences. Nonetheless, starting from first principles to obtain continuous quantum master equations involves various approximations such as weak coupling to the environment. Further, converting these equations to Kraus operators cannot generally be obtained in closed-form due to the complicated commutator structure of the problem. In our work, we bridge this gap by providing a general closed form formulation for arbitrarily strong driving while remaining linear in the dissipator. The Kraus solution is expressed as a Riemann sum where higher terms can converge quickly to high precision, which we demonstrate numerically. Such a formulation is highly relevant to quantum computing and gate-based models, where effective models are highly sought for large rotation gate angles, even under the influence of underlying non-trivial noise mechanisms.
\end{abstract}

\maketitle

Quantum mechanics is generally studied in two different formulations: dynamical, describing time evolution \cite{breuer2002theory}, and informational, describing coarse-grained content through a quantum channel \cite{TQI}.
While the former is fundamental to understanding the diffusion physics of probability amplitudes, the latter provides a necessary abstraction for describing the discretized form of quantum information manipulation, especially under the influence of quantum measurement. 

A Kraus-Lindbladian correspondence acts like a dictionary to map the two languages, of open-quantum dynamics and of quantum information. 
A straightforward numerical approach is to diagonalize the Choi matrix of the corresponding quantum map, retrieving eigenvectors as canonical Kraus operators \cite{robust-conversions-QDS, Andersson15082007}.
Open dynamics can also be solved with integration methods such as Runge-Kutta, and then similarly diagonalized to identify Kraus operators \cite{KiK}. The major drawback with the numerical approach however is the time-consuming requirement to recalculate them for every problem instance.

Analytically, first order approximations of the Liouvillian generator in time have been applied to obtain a direct Kraus-Lindbladian correspondence \cite{breuer2002theory, ERT}.
These inherently require that the generator of the maps be small, which is to say that the product of energies and dissipation rates with evolution time is minimized. Nonetheless, such techniques have provided a trough of useful results. Time ordered approximation of the dynamics can further be extended to high-order unraveling \cite{unraveling}, as already achieved in \cite{unraveling-averageKraus,hamiltonian-circuit-simulation}.
Of particular note, for specific problems, Operator Sum Representation (OSR) methods identify the spectrum of system-bath Hamiltonian and trace-out the bath avoiding the master equation formalism \cite{d-level-thermal-bath, damped-Harmonic-oscillator}. 
Interestingly, OSR could also be manipulated in some cases to yield semi-group master equations (GKSL), without tracing the bath sub-system \cite{robustness-DF-subspace}.
Moreover, there exists always an OSR connection between two density matrices \cite{OSR-density-matrix}. 
For a special case of commuting super-operators of unitary and dissipative generators, an exact closed-form solution of the Kraus operators could be derived, solving the coupled differential equations \cite{commuting-super-exact-Kraus}. 

A generalized closed-form solution however has remained elusive. In most cases, the first order approximation has only been possible, thereby completely understating the role of non-commuting components in the dynamics.

In this \paper, we derive Kraus operators in a closed-form, eliminating the need for weakly driven coherent dynamics while retaining the assumption of weak dissipation. This result is general over weak Markovian noise channels and ranks of the corresponding jump operators. 
To identify the Kraus operators, we evaluate dynamics in the superoperator representation, and segregate coherent, decoherent  terms and their interaction. Consequently, the closed-form solution is described in the eigenbasis of the coherent map. 
Because of the linearization, the derived Kraus operators as a result are not trace-preserving. 
To be concise on the description of strategy, we consider a finite-dimensional($d$) time-static Hamiltonian. However, the problem with a time-dependent Hamiltonian could be similarly solved, having it evolved under time ordered exponential, for example via Magnus expansion \cite{Magnus-Expansion}.

While the remaining assumption of weak dissipation can be seen as a strong restriction, it can also be seen as consistent with standard requirements in both technological and fundamental physics application. That is, standard GKSL derivations operate under conditions such as Born, Markov, and secular approximations already greatly constraining any physical interpretation to effective Kraus maps. At the technological level, it is of prime importance to understand physically the role of different error mechanisms on quantum information processing capabilities. Therefore, while dissipation is engineered to play a minimal role for logical operations, it is crucial to understand what the effective channel under such perturbation is, even and especially when the coherent dynamics are themselves prominent.

\section{Open quantum system dynamics}\label{sec:open dynamics}
The dynamics of an open quantum system interacting with its environment under the Born-Markov approximation is governed by the GKSL master equation~\cite{breuer2002theory}
\begin{equation}\label{def:GKSL}
    \dot{\rho} = - \im [H,\rho] + \sum_\l \gamma_\l \bigg[ { L_\l \rho \conj{L_\l}- \frac{1}{2}\{\conj{L_\l} L_\l, \rho \}_+} \bigg]
\end{equation}
where $\rho(t)$ is the state of the quantum system evolving over $0 \leq t \leq \tau$ , $L_\l$ is the collapse operator describing dissipation at rate $\gamma_\l$, for each noise channel. For simplification, we consider a time-independent system Hamiltonian ($H$).

Vectorization of the master equation (Eq.~\ref{def:GKSL}) yields the superoperator Liouvillian($\L$)  equation 
$
    \dot{\vec{\rho}} = \L \vec{\rho} 
$.
Exponentiating its integral results in the natural representation  ($K_\Map$) of the quantum dynamical map ($\Map$) \cite{TQI}  with 
\begin{equation}\label{eq:liouvillian}
       \vec{\rho_\tau} = e^{\int^\tau \L}  \vec{\rho_0} \equiv K_\Map \vec{\rho_0}.
\end{equation}

The Kraus theorem \cite{kraus-theorem} establishes that a CPTP quantum map can be expressed with a set of operators $\{K_k\}$ such that $\Map(\rho) = \sum_k K_k \rho \conj{K_k}, \quad \sum_k \conj{K_k} K_k  = \I$.
In equivalent terms, vectorization of the Kraus representation yields the natural representation as a decomposition -- $ K_\Map = \sum_k^N K_k^* \otimes K_k$ -- acting in the Hilbert-Schmidt space.

\section{Linear order dissipation dynamics}
We carry out a linearization of the master equation in the superoperator space in terms of the dissipation. 

First, a Zassenhaus expansion of the Liouvillian exponential is applied, restricted to linear order nesting in noise, assuming a dominance of  coherent dynamics.
Weak dissipation, such that $\gamma \tau \ll |\lambda| \tau$, is inherited from the secular approximation, $\gamma \ll |\Delta\lambda| < |\lambda|$ in the GKSL formulation, where $\lambda$ describes the (eigen)energies of the system.

Second, for a given evolution time, we have a linearization in  
$ \tau \gamma_\l  \ll 1 $, whereby dissipative exponentials in the previous expansion can be expanded, segregating different noise channels.
Note that these are significantly weaker assumptions than are usually enforced for Kraus operator derivations, and we place no limitation here on $\lambda$. 

Now, we break down the Liouvillian super-operator of the master equation \eqref{def:GKSL} in terms of coherent ($\C$) and dissipative ($D+\Dc$) dynamics generators with 
\begin{eqnarray}\label{bch:1order}
    \L &=& \C - \Dc + D \\
    \C &=& (-\im H\tau)^* \oplus (-\im H\tau) \nonumber\\ 
    \Dc& =& \frac{1}{2} \big[ (\err_\l \conj{L_\l}L_\l)^* \oplus (\err_\l \conj{L_\l}L_\l) \big]  \nonumber\\
     C &=& \C - \Dc \quad D = \err_\l L^*_\l\otimes L_\l, \nonumber   
\end{eqnarray} in terms of kronecker product ($\otimes$) and sum ($\oplus$) with summation over repeated indices of noise channels ($\l$).

Combining the above two simplifications, a linearized dynamical map ($\MapI$) can be derived as
\begin{equation}
    K_{\MapI} = e^C \bigg(\I + \sum_{n=0}^\infty \frac{[\C,D]_n}{(n+1)! }+\O(\err^2) \bigg), 
\end{equation}
where $C$ and $D$ are the rearrangement of generators into Kraus-conforming and non-conforming form, respectively, while $\err\equiv\gamma\tau$ is the small parameter.   

The nested commutator structure can be reduced by moving to the eigenbasis of the coherent generator, defined as $\C\ket{i}=-\im \mu_i\ket{i}$.
Eigenstates are vectorized in terms of the system Hamiltonian so that $\ket{i}=\ket{i_a}^*\ket{i_b}$, where $\mu_i = \lambda_{i_a}^* - \lambda_{i_b}$ with $H^* \ket{i_a}^*=\eig{i_a}^*\ket{i_a}^*$, $H \ket{i_b}=\eig{i_b}\ket{i_b}$.

Then, we see that $\bra{i}[\C,D]_n\ket{j} = (-\im\Delta\mu_{ij})^n D_{ij}$ with $D_{ij} \coloneqq \bra{i}D\ket{j}$ and $\Delta\mu_{ij}  \equiv \mu_i^* - \mu_j$, which dresses the collapse operators.  
Each term $D_{ij}$ appears to be a coefficient of linearized dynamical friction.
This results in a closed-form representation 
\begin{equation}\label{1st order}
     K_{\MapI} = e^C\bigg(\I + \sum_{ij} (D \odot R)_{ij} \ket{i}\bra{j}\bigg),
 \end{equation} 
with element-wise (Hadamard) product defined in the eigenbasis projector ($\ket{i}\bra{j}$) and interaction factor  $R_{ij} \equiv (1-e^{-\im \Delta\mu_{ij}})/(\im \Delta\mu_{ij}) $. The interaction factor appears similarly in the Redfield equation as well \cite{breuer2002theory}.

This provides a closed form for the effective dynamics but it is not yet in the Kraus form.

\section{Riemann Summation}
In Eq.~\eqref{1st order}, individually, dressing ($e^C$) \& dissipative ($D$) terms  admit a Kraus decomposition [via Eq.~\eqref{bch:1order}].
However, their interaction  
fails such a representation.

Now, we introduce a second approximation to the interaction factor, noticing that the interaction factor can be expressed as the result of an integral, with the latter computed through Riemann summation:
\begin{gather}\label{Riemann-sum}
    R_{ij} = \int_0^1 dt \ e^{-\im (\mu_i- \mu_j) t} =
    \lim_{N \to \infty} \sum_k^N \Delta_k{(e^{-\im (\eig{i_a}-\eig{j_a})\tau_k})}^* e^{-\im (\eig{i_b}-\eig{j_b})\tau_k},
\end{gather}
where we express the superoperator $R$ in terms of pair of transition spectrum of the Hamiltonian.

The trick here is that while the coefficient $\bra{i}R\ket{j} = R_{ij}=(1-e^{-\im \Delta\mu})/\im \Delta\mu$ could not be factorized analytically (in $a,b$ projection), individual samples from Eq.~\eqref{Riemann-sum} decompose for all $(i,j)$ pairs, similar to $D_{ij}$ in multi-index ($i_a,i_b,j_a,j_b$) notation.- 
Thus we have the factorization
\begin{eqnarray}
    R &=& \sum_k^N {R^{(k)}{}}^* \otimes \R{k}{} \\
     {( {\R{k}{}}^* \otimes\R{k}{} )}_{ij} &=& \bra{i_a}^* \bra{i_b}
       {R^{(k)}{}}^* \otimes \R{k}{} \ket{j_a}^* \ket{j_b} \nonumber\\
      &=& ({\bra{i_a}\R{k}{}\ket{j_a}})^* \bra{i_b}\R{k}{}\ket{j_b} \nonumber\\
     &=&{(e^{-\im (\eig{i_a}-\eig{j_a}))\tau_k})}^* e^{-\im (\eig{i_b}-\eig{j_b})\tau_k}{\sqrt{\Delta_k}\sqrt{\Delta_k}},\nonumber
\end{eqnarray} which approximates the nested commutators in Kraus-type representation :
\begin{gather*}
    (D \odot R)_{ij} = \sum_\l \sum_k^{N_\l} \gamma_\ell \bra{i}(L^*_\l \otimes L_\l)\odot ({\R{k}{}}^* \otimes \R{k}{})\ket{j} \\
    = \sum_\l \sum_k^{N_\l} \gamma_\ell \bra{i}{(L_\l\odot \R{k}{})}^*\otimes (L_\l \odot \R{k}{})\ket{j},
\end{gather*}
 where each noise channel ($\l$) can exhibit independent Riemann summation ($N_\l$).

The approximation truncates the summation to a finite number of quadratures ($N$). Note that any alternate design of quadrature should also be sufficient to derive the Kraus operators.

Finally, Kraus operators for the approximate map $\MapN$, in the coherent basis are identified in the closed-form  
\begin{eqnarray}
    \label{kraus-operators}
    \K{0}{} &=& e^{-\im H-\sum_\l \gamma_\ell \conj{L_\l}L_\l/2} \\ 
    \K{k}{\l} &=& e^{-\im H-\sum_\l \gamma_\ell \conj{L_\l}L_\l/2} \sum_{m,n=1}^d (L_l \odot \R{k}{})_{mn} \ket{m}\bra{n} , \ k \leq  N_l. \nonumber
\end{eqnarray}

The weak noise approximation ($\err \ll 1 $) results in $\norm{K_0} \sim 1$ describing almost coherent dynamics, and interaction terms as corrections $\norm{\K{k}{\l}} \sim \err_l$.
The description of correction terms ($\K{k}{\l}$) in the coherent basis represents the dressing of interaction under frame transformation ($\K{0}{}$).

\section{Radius of Convergence}\label{sec:RoC}
For simplicity, consider a uniform width ($ \Delta_k=1/N$) with mid-point sampling ($\tau_k = (2k-1)\Delta/2 , \ k \in \{1,2,\cdots N \}$) for the quadrature design. 
A linear order approximation ($\MapI$) does not preserve trace (non-TP), and is justified only for limited time \cite{CPTP-long-time} --- $\tau \ll \inv{\gamma}$.
Building upon them, the Kraus approximation (\ref{kraus-operators}) incurs additional error in the Riemann resummation of interaction terms. For mid-point sampling, we have that $\norm{R_{ij}-\sum_k {{(\R{k}{}}^* \otimes \R{k}{})}_{ij}} \sim \O(1/ N_\l^2)$.
Therefore, the error in Kraus closure scales as $\sum_k^N \sum_\l \conj{\K{k}{\l}} \K{k}{\l} = \I + \O(\sum_\l \err_\l / N_\l^2) + \O(\err^2)$, and
is regulated by the quadrature approximation.

Including the interaction of the dissipator with coherent dynamics improves the validity of long evolution time ($\tau$) when compared to the infinitesimal map ($d\Map$) in earlier methods \cite{breuer2002theory}.
For example, infinitesimal evolution of the master equation results in a direct Kraus-Lindblad correspondence. However, the limit of this timescale ($\tau_{d\Map}^*$) is defined by the inverse of the largest relevant energy scale $\lambda$ in the system. For instance, without the rotating wave approximation this energy will be the bare frequency of the system, meaning that the evolution will barely capture a fraction of a cycle before it starts to break down. Similarly, if a Rabi frequency is relevant, then once again the modeled dynamics can only capture a fraction of a Rabi oscillation, meaning most gates cannot be accurately modeled. The minimum time thus grows inversely proportional to the relevant energy, or $\tau^*_{d\Map}\sim\sqrt{\err_0}/\lambda$, where $\err_0$ is the tolerance of error.

On the other hand, our linear order approximation ($\Map^{(\err)}$) also incurs a quadratic error, but now in noise rate, 
$\tau^*_{\Map^{(\err)}}
\sim \sqrt{\err_0}/\gamma$.
As a requirement on the master equation formulation as well as any technological application, the transition energy  is assumed to be significantly stronger than noise, i.e.~$\lambda \gg \gamma$. 
At its saturation, it results in a longer available convergence time of $\Map^{(\err)}$  so that $\tau^*_{d\Map} \ll  \tau^*_{\Map^{(\err)}}$.

 The Riemann approximated linear map ($ \Map^{(\err)}_N $) builds additional error on the linear approximation ($ \err = \gamma \tau $), regulated by the number of quadratures($ N $), and is $ \O(\err/N^2) + \O(\err^2) < \err_0 $. This places the radius of convergence for Kraus approximated map in between, $\tau^*_{d\Map} \ll \tau^*_{\Map^{(\err)}_N} < \tau^*_{\Map^{(\err)}}$, approaching the linearized map for higher number of quadratures. Next, we study numerically typical requirements for $N$.

\section{Example}\label{sec:example}
We present a numerical example to compare the accuracy of the proposed closed-form solution and various approximations. 
We choose a fictitious noise model with high rank collapse operators to highlight the efficacy over exact low-rank Lindbladians.

Consider a 3-level system implementing a single qubit gate, ubiquitous in optimal control development. Without the loss of generality, its Hamiltonian under RWA of the drive can be described as 
\begin{gather} \label{def:hamiltonian-3LS}
    H = \begin{bmatrix}
        0 & \Omega & 0 \\ \Omega & 0 & \Omega \eta \\ 0 & \Omega \eta & - \alpha
    \end{bmatrix},
\end{gather} with a $\ket{1} \leftrightarrow \ket{2}$ leakage strength $\eta$, $\alpha$ as anharmonicity, and $\Omega$ as time-constant control strength.

The open-system dynamics is simulated under a single noise channel described by the Lindbladian  
\begin{gather} \label{def:lind-3LS}
    L =\ket{0}\bra{0}+\ket{1}\bra{0}+\ket{1}\bra{1}+\ket{1}\bra{2},
\end{gather} that could be interpreted as a combination of dephasing and relaxation noise.

We simulate the GKSL dynamics of this system in the superoperator space. A straightforward exponentiation (Eq.~\ref{eq:liouvillian}) yields the exact dynamical map $K_\Map$. The first order simulation is calculated with closed-form representation in Eq.~\ref{1st order} as $K_\MapI$ and its Kraus-approximation for different ranks ($N=1,10,50$) are evaluated with the proposed solution (Eq.\ref{kraus-operators}).
\begin{figure}[H]
    \centering
    \includegraphics[width=\columnwidth]{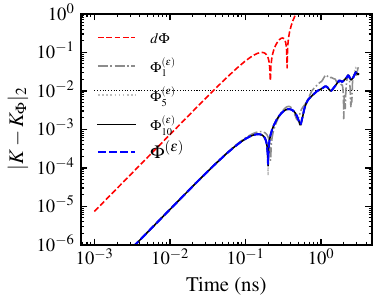}
    \caption{Comparison of accuracy of super-operator representation of various \Ist order approximations over total time of evolution (x-axis). The accuracy is calculated over $L_2$ norm against the difference of exact (exponential) evolution. $d\Map$ is infinitesimal time evolution of the Liouvillian. Small error regulator is fixed at $\err_0 = .01$ (horizontal dotted). Simulation parameters for Hamiltonian with leakage in Eq.~\eqref{def:hamiltonian-3LS}: $ \Omega=\pi, \ \alpha = \Omega/20 , \ \eta = \sqrt{2}$ with a decay rate $\gamma = \err_0 \Omega$.}
    \label{fig:error-L2-N}
\end{figure}
These approximations are compared with the exact result ($K_\Map$) in the L2 norm of their difference  (Fig.~\ref{fig:error-L2-N}).
We find accuracy of Kraus-approximation maps ($\MapI_1,\MapI_{10},\MapI_{50}$) at par with the first order approximation ($\MapI$).
Higher quadrature resummation ($N=10,50$) yields long-time agreement compared to the single corrector Kraus operator ($\MapI_{1}$).

The radius of convergence for the first order map ($\MapI$) is defined as $ \tau \lesssim \tau_* \equiv \sqrt{\err_0} / {\gamma} \sim 3ns$. Within this period of approximation, the proposed model (Eq.~\ref{def:hamiltonian-3LS}) performs the X-gate in $1ns$. Its noisy dynamics is effectively captured with the proposed Kraus-approximation .

\begin{figure}[H]
    \centering
    \includegraphics[width=\columnwidth]{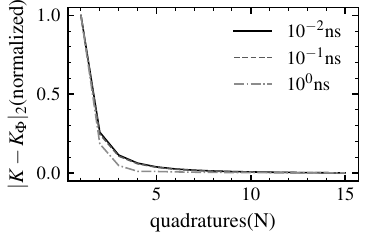}
    \caption{Increasing accuracy of Kraus-approximation maps with number of quadratures (x-axis) describing the interaction ($R$). Accuracy comparison against exact simulation is normalized for three different times of simulation ($\tau = 1,.1,.01$ns). Simulation parameters are same as in Fig.~\ref{fig:error-L2-times}.}
    \label{fig:error-L2-times}
\end{figure}

Over different duration times of simulation under the limit of convergence ($\tau < \tau_* $), we identify that the numerical accuracy of Kraus-approximations increasing with the number of quadratures (Fig.~\ref{fig:error-L2-times}).

\section{Discussion \& Summary}\label{sec:conclusion}
In this \paper, we have derived a closed-form solution to Kraus operators of a GKSL master equation. The closed-form is described in the basis of coherent dynamics for any general noise model.

Our method directly benefits from the approximation on noise strength, instead of short-time expansion on the linearized dynamics.
The numerical example confirms this improvement over the latter method, resulting in a larger radius of convergence.
These solution establish a direct correspondence between Kraus operators and Lindbladians.

A closed-form solution of the Kraus-operators, albeit an approximation, provides a physical picture of the open-system dynamics.  
We identify a coherent dominating and interaction-correction structure in Kraus-operators representation. In this format, Riemann sampling bundles the time averaged interaction of coherent dynamics with dissipators of various noise channels, independently.

From a computational perspective, our calculation result in a closed-form solution to the eigenvalue decomposition of a Choi matrix, that could be made arbitrarily accurate.
The Riemann trick employed here is similar to the Nyström method, where a quadrature method approximate the eigenvalue decomposition of a Gram matrix \cite{Nystrom-Gram}.

Application of our closed-form solution could benefit ansatz design of open-system dynamics. For example, in process tomography (QPT), arbitrarily ranked noise channels could be efficiently instantiated. Faster algorithms could be built upon, that require less observations, like improvements to \cite{GD-QPT}.
Additionally, it motivates a phenomenological hypothesis to begin with.
In a similar way, error mitigation techniques and modeling of weak continuous measurement could be upgraded.

Further, memory utilization of numerical computation with Hilbert space scaling could be improved quadratically with Kraus representation compared to the natural representation of quantum dynamical maps.

For control problems, time-ordered integration could be handled via Magnus approximation to render the coherent and dissipative generators in the vectorized representation.

In principle, it could be extended to cover CP-divisible non-Markovian models as well. The GKSL in that case would be upgraded with time-dependent decay rates($\gamma(t)$) \cite{breuer2002theory}, keeping the tensor structure intact upon integration over time.

On a more abstract outlook, we believe a general theory on deriving Kraus operators from quadrature sampling could be motivated with this example.

\section{Acknowdedgements}
This work was funded by the Federal Ministry of Education and Research (BMBF) within the framework programme "Quantum technologies - from basic research to market" (Project QSolid, Grant No. 13N16149) and by Horizon Europe program via project QCFD (101080085, HORIZON-CL4-2021-DIGITAL-EMERGING02-10), project OpenSuperQPlus100 (101113946, HORIZON-CL4-2022-QUANTUM-01-SGA) and by the Cluster of Excellence Matter and Light for Quantum Computing (ML4Q2) EXC 2004/1 – 390534769.


\begin{thebibliography}{19}%
\makeatletter
\providecommand \@ifxundefined [1]{%
 \@ifx{#1\undefined}
}%
\providecommand \@ifnum [1]{%
 \ifnum #1\expandafter \@firstoftwo
 \else \expandafter \@secondoftwo
 \fi
}%
\providecommand \@ifx [1]{%
 \ifx #1\expandafter \@firstoftwo
 \else \expandafter \@secondoftwo
 \fi
}%
\providecommand \natexlab [1]{#1}%
\providecommand \enquote  [1]{``#1''}%
\providecommand \bibnamefont  [1]{#1}%
\providecommand \bibfnamefont [1]{#1}%
\providecommand \citenamefont [1]{#1}%
\providecommand \href@noop [0]{\@secondoftwo}%
\providecommand \href [0]{\begingroup \@sanitize@url \@href}%
\providecommand \@href[1]{\@@startlink{#1}\@@href}%
\providecommand \@@href[1]{\endgroup#1\@@endlink}%
\providecommand \@sanitize@url [0]{\catcode `\\12\catcode `\$12\catcode
  `\&12\catcode `\#12\catcode `\^12\catcode `\_12\catcode `\%12\relax}%
\providecommand \@@startlink[1]{}%
\providecommand \@@endlink[0]{}%
\providecommand \url  [0]{\begingroup\@sanitize@url \@url }%
\providecommand \@url [1]{\endgroup\@href {#1}{\urlprefix }}%
\providecommand \urlprefix  [0]{URL }%
\providecommand \Eprint [0]{\href }%
\providecommand \doibase [0]{https://doi.org/}%
\providecommand \selectlanguage [0]{\@gobble}%
\providecommand \bibinfo  [0]{\@secondoftwo}%
\providecommand \bibfield  [0]{\@secondoftwo}%
\providecommand \translation [1]{[#1]}%
\providecommand \BibitemOpen [0]{}%
\providecommand \bibitemStop [0]{}%
\providecommand \bibitemNoStop [0]{.\EOS\space}%
\providecommand \EOS [0]{\spacefactor3000\relax}%
\providecommand \BibitemShut  [1]{\csname bibitem#1\endcsname}%
\let\auto@bib@innerbib\@empty
\bibitem [{\citenamefont {Breuer}\ and\ \citenamefont
  {Petruccione}(2002)}]{breuer2002theory}%
  \BibitemOpen
  \bibfield  {author} {\bibinfo {author} {\bibfnamefont {H.}~\bibnamefont
  {Breuer}}\ and\ \bibinfo {author} {\bibfnamefont {F.}~\bibnamefont
  {Petruccione}},\ }\href {https://books.google.de/books?id=0Yx5VzaMYm8C}
  {\emph {\bibinfo {title} {The Theory of Open Quantum Systems}}}\ (\bibinfo
  {publisher} {Oxford University Press},\ \bibinfo {year} {2002})\BibitemShut
  {NoStop}%
\bibitem [{\citenamefont {Watrous}(2018)}]{TQI}%
  \BibitemOpen
  \bibfield  {author} {\bibinfo {author} {\bibfnamefont {J.}~\bibnamefont
  {Watrous}},\ }\href@noop {} {\emph {\bibinfo {title} {The Theory of Quantum
  Information}}},\ \bibinfo {edition} {1st}\ ed.\ (\bibinfo  {publisher}
  {Cambridge University Press},\ \bibinfo {address} {USA},\ \bibinfo {year}
  {2018})\BibitemShut {NoStop}%
\bibitem [{\citenamefont {Havel}(2003)}]{robust-conversions-QDS}%
  \BibitemOpen
  \bibfield  {author} {\bibinfo {author} {\bibfnamefont {T.~F.}\ \bibnamefont
  {Havel}},\ }\bibfield  {title} {\bibinfo {title} {Robust procedures for
  converting among lindblad, kraus and matrix representations of quantum
  dynamical semigroups},\ }\href {https://doi.org/10.1063/1.1518555} {\bibfield
   {journal} {\bibinfo  {journal} {Journal of Mathematical Physics}\ }\textbf
  {\bibinfo {volume} {44}},\ \bibinfo {pages} {534} (\bibinfo {year} {2003})},\
  \BibitemShut {NoStop}%
\bibitem [{\citenamefont {Andersson}\ \emph {et~al.}(2007)\citenamefont
  {Andersson}, \citenamefont {Cresser},\ and\ \citenamefont
  {Hall}}]{Andersson15082007}%
  \BibitemOpen
  \bibfield  {author} {\bibinfo {author} {\bibfnamefont {E.}~\bibnamefont
  {Andersson}}, \bibinfo {author} {\bibfnamefont {J.~D.}\ \bibnamefont
  {Cresser}},\ and\ \bibinfo {author} {\bibfnamefont {M.~J.~W.}\ \bibnamefont
  {Hall}},\ }\bibfield  {title} {\bibinfo {title} {Finding the kraus
  decomposition from a master equation and vice versa},\ }\href
  {https://doi.org/10.1080/09500340701352581} {\bibfield  {journal} {\bibinfo
  {journal} {Journal of Modern Optics}\ }\textbf {\bibinfo {volume} {54}},\
  \bibinfo {pages} {1695} (\bibinfo {year} {2007})},\ 
\bibitem [{\citenamefont {Appelö}\ and\ \citenamefont {Cheng}(2025)}]{KiK}%
  \BibitemOpen
  \bibfield  {author} {\bibinfo {author} {\bibfnamefont {D.}~\bibnamefont
  {Appelö}}\ and\ \bibinfo {author} {\bibfnamefont {Y.}~\bibnamefont
  {Cheng}},\ }\bibfield  {title} {\bibinfo {title} {Kraus is king: High-order
  completely positive and trace preserving (cptp) low rank method for the
  lindblad master equation},\ }\href
  {https://doi.org/https://doi.org/10.1016/j.jcp.2025.114036} {\bibfield
  {journal} {\bibinfo  {journal} {Journal of Computational Physics}\ }\textbf
  {\bibinfo {volume} {534}},\ \bibinfo {pages} {114036} (\bibinfo {year}
  {2025})}\BibitemShut {NoStop}%
\bibitem [{\citenamefont {McCaul}\ \emph {et~al.}(2021)\citenamefont {McCaul},
  \citenamefont {Jacobs},\ and\ \citenamefont {Bondar}}]{ERT}%
  \BibitemOpen
  \bibfield  {author} {\bibinfo {author} {\bibfnamefont {G.}~\bibnamefont
  {McCaul}}, \bibinfo {author} {\bibfnamefont {K.}~\bibnamefont {Jacobs}},\
  and\ \bibinfo {author} {\bibfnamefont {D.~I.}\ \bibnamefont {Bondar}},\
  }\bibfield  {title} {\bibinfo {title} {Fast computation of dissipative
  quantum systems with ensemble rank truncation},\ }\href
  {https://doi.org/10.1103/PhysRevResearch.3.013017} {\bibfield  {journal}
  {\bibinfo  {journal} {Phys. Rev. Res.}\ }\textbf {\bibinfo {volume} {3}},\
  \bibinfo {pages} {013017} (\bibinfo {year} {2021})}\BibitemShut {NoStop}%
\bibitem [{\citenamefont {Steinbach}\ \emph {et~al.}(1995)\citenamefont
  {Steinbach}, \citenamefont {Garraway},\ and\ \citenamefont
  {Knight}}]{unraveling}%
  \BibitemOpen
  \bibfield  {author} {\bibinfo {author} {\bibfnamefont {J.}~\bibnamefont
  {Steinbach}}, \bibinfo {author} {\bibfnamefont {B.~M.}\ \bibnamefont
  {Garraway}},\ and\ \bibinfo {author} {\bibfnamefont {P.~L.}\ \bibnamefont
  {Knight}},\ }\bibfield  {title} {\bibinfo {title} {High-order unraveling of
  master equations for dissipative evolution},\ }\href
  {https://doi.org/10.1103/PhysRevA.51.3302} {\bibfield  {journal} {\bibinfo
  {journal} {Phys. Rev. A}\ }\textbf {\bibinfo {volume} {51}},\ \bibinfo
  {pages} {3302} (\bibinfo {year} {1995})}\BibitemShut {NoStop}%
\bibitem [{\citenamefont {Wonglakhon}\ \emph {et~al.}(2024)\citenamefont
  {Wonglakhon}, \citenamefont {Wiseman},\ and\ \citenamefont
  {Chantasri}}]{unraveling-averageKraus}%
  \BibitemOpen
  \bibfield  {author} {\bibinfo {author} {\bibfnamefont {N.}~\bibnamefont
  {Wonglakhon}}, \bibinfo {author} {\bibfnamefont {H.~M.}\ \bibnamefont
  {Wiseman}},\ and\ \bibinfo {author} {\bibfnamefont {A.}~\bibnamefont
  {Chantasri}},\ }\bibfield  {title} {\bibinfo {title} {Completely positive
  trace-preserving maps for higher-order unraveling of lindblad master
  equations},\ }\href {https://doi.org/10.1103/PhysRevA.110.062207} {\bibfield
  {journal} {\bibinfo  {journal} {Phys. Rev. A}\ }\textbf {\bibinfo {volume}
  {110}},\ \bibinfo {pages} {062207} (\bibinfo {year} {2024})}\BibitemShut
  {NoStop}%
\bibitem [{\citenamefont {Ding}\ \emph {et~al.}(2024)\citenamefont {Ding},
  \citenamefont {Li},\ and\ \citenamefont
  {Lin}}]{hamiltonian-circuit-simulation}%
  \BibitemOpen
  \bibfield  {author} {\bibinfo {author} {\bibfnamefont {Z.}~\bibnamefont
  {Ding}}, \bibinfo {author} {\bibfnamefont {X.}~\bibnamefont {Li}},\ and\
  \bibinfo {author} {\bibfnamefont {L.}~\bibnamefont {Lin}},\ }\bibfield
  {title} {\bibinfo {title} {Simulating open quantum systems using hamiltonian
  simulations},\ }\href {https://doi.org/10.1103/PRXQuantum.5.020332}
  {\bibfield  {journal} {\bibinfo  {journal} {PRX Quantum}\ }\textbf {\bibinfo
  {volume} {5}},\ \bibinfo {pages} {020332} (\bibinfo {year}
  {2024})}\BibitemShut {NoStop}%
\bibitem [{\citenamefont {Biswas}\ and\ \citenamefont
  {Brumer}(2012)}]{d-level-thermal-bath}%
  \BibitemOpen
  \bibfield  {author} {\bibinfo {author} {\bibfnamefont {A.}~\bibnamefont
  {Biswas}}\ and\ \bibinfo {author} {\bibfnamefont {P.}~\bibnamefont
  {Brumer}},\ }\bibfield  {title} {\bibinfo {title} {Generic construction of
  kraus operators: d-level systems in a thermal bosonic bath},\ }\href
  {https://doi.org/https://doi.org/10.1002/ijch.201100114} {\bibfield
  {journal} {\bibinfo  {journal} {Israel Journal of Chemistry}\ }\textbf
  {\bibinfo {volume} {52}},\ \bibinfo {pages} {461} (\bibinfo {year} {2012})},\
  \Eprint
  {https://arxiv.org/abs/https://onlinelibrary.wiley.com/doi/pdf/10.1002/ijch.201100114}
  {https://onlinelibrary.wiley.com/doi/pdf/10.1002/ijch.201100114} \BibitemShut
  {NoStop}%
\bibitem [{\citenamefont {Liu}\ \emph {et~al.}(2004)\citenamefont {Liu},
  \citenamefont {\"Ozdemir}, \citenamefont {Miranowicz},\ and\ \citenamefont
  {Imoto}}]{damped-Harmonic-oscillator}%
  \BibitemOpen
  \bibfield  {author} {\bibinfo {author} {\bibfnamefont {Y.-x.}\ \bibnamefont
  {Liu}}, \bibinfo {author} {\bibfnamefont {i.~m. c.~K.}\ \bibnamefont
  {\"Ozdemir}}, \bibinfo {author} {\bibfnamefont {A.}~\bibnamefont
  {Miranowicz}},\ and\ \bibinfo {author} {\bibfnamefont {N.}~\bibnamefont
  {Imoto}},\ }\bibfield  {title} {\bibinfo {title} {Kraus representation of a
  damped harmonic oscillator and its application},\ }\href
  {https://doi.org/10.1103/PhysRevA.70.042308} {\bibfield  {journal} {\bibinfo
  {journal} {Phys. Rev. A}\ }\textbf {\bibinfo {volume} {70}},\ \bibinfo
  {pages} {042308} (\bibinfo {year} {2004})}\BibitemShut {NoStop}%
\bibitem [{\citenamefont {Bacon}\ \emph {et~al.}(1999)\citenamefont {Bacon},
  \citenamefont {Lidar},\ and\ \citenamefont
  {Whaley}}]{robustness-DF-subspace}%
  \BibitemOpen
  \bibfield  {author} {\bibinfo {author} {\bibfnamefont {D.}~\bibnamefont
  {Bacon}}, \bibinfo {author} {\bibfnamefont {D.~A.}\ \bibnamefont {Lidar}},\
  and\ \bibinfo {author} {\bibfnamefont {K.~B.}\ \bibnamefont {Whaley}},\
  }\bibfield  {title} {\bibinfo {title} {Robustness of decoherence-free
  subspaces for quantum computation},\ }\href
  {https://doi.org/10.1103/PhysRevA.60.1944} {\bibfield  {journal} {\bibinfo
  {journal} {Phys. Rev. A}\ }\textbf {\bibinfo {volume} {60}},\ \bibinfo
  {pages} {1944} (\bibinfo {year} {1999})}\BibitemShut {NoStop}%
\bibitem [{\citenamefont {Tong}\ \emph {et~al.}(2004)\citenamefont {Tong},
  \citenamefont {Kwek}, \citenamefont {Oh}, \citenamefont {Chen},\ and\
  \citenamefont {Ma}}]{OSR-density-matrix}%
  \BibitemOpen
  \bibfield  {author} {\bibinfo {author} {\bibfnamefont {D.~M.}\ \bibnamefont
  {Tong}}, \bibinfo {author} {\bibfnamefont {L.~C.}\ \bibnamefont {Kwek}},
  \bibinfo {author} {\bibfnamefont {C.~H.}\ \bibnamefont {Oh}}, \bibinfo
  {author} {\bibfnamefont {J.-L.}\ \bibnamefont {Chen}},\ and\ \bibinfo
  {author} {\bibfnamefont {L.}~\bibnamefont {Ma}},\ }\bibfield  {title}
  {\bibinfo {title} {Operator-sum representation of time-dependent density
  operators and its applications},\ }\href
  {https://doi.org/10.1103/PhysRevA.69.054102} {\bibfield  {journal} {\bibinfo
  {journal} {Phys. Rev. A}\ }\textbf {\bibinfo {volume} {69}},\ \bibinfo
  {pages} {054102} (\bibinfo {year} {2004})}\BibitemShut {NoStop}%
\bibitem [{\citenamefont {Nakazato}\ \emph {et~al.}(2006)\citenamefont
  {Nakazato}, \citenamefont {Hida}, \citenamefont {Yuasa}, \citenamefont
  {Militello}, \citenamefont {Napoli},\ and\ \citenamefont
  {Messina}}]{commuting-super-exact-Kraus}%
  \BibitemOpen
  \bibfield  {author} {\bibinfo {author} {\bibfnamefont {H.}~\bibnamefont
  {Nakazato}}, \bibinfo {author} {\bibfnamefont {Y.}~\bibnamefont {Hida}},
  \bibinfo {author} {\bibfnamefont {K.}~\bibnamefont {Yuasa}}, \bibinfo
  {author} {\bibfnamefont {B.}~\bibnamefont {Militello}}, \bibinfo {author}
  {\bibfnamefont {A.}~\bibnamefont {Napoli}},\ and\ \bibinfo {author}
  {\bibfnamefont {A.}~\bibnamefont {Messina}},\ }\bibfield  {title} {\bibinfo
  {title} {Solution of the lindblad equation in the kraus representation},\
  }\href {https://doi.org/10.1103/PhysRevA.74.062113} {\bibfield  {journal}
  {\bibinfo  {journal} {Phys. Rev. A}\ }\textbf {\bibinfo {volume} {74}},\
  \bibinfo {pages} {062113} (\bibinfo {year} {2006})}\BibitemShut {NoStop}%
\bibitem [{\citenamefont {Blanes}\ \emph {et~al.}(2009)\citenamefont {Blanes},
  \citenamefont {Casas}, \citenamefont {Oteo},\ and\ \citenamefont
  {Ros}}]{Magnus-Expansion}%
  \BibitemOpen
  \bibfield  {author} {\bibinfo {author} {\bibfnamefont {S.}~\bibnamefont
  {Blanes}}, \bibinfo {author} {\bibfnamefont {F.}~\bibnamefont {Casas}},
  \bibinfo {author} {\bibfnamefont {J.}~\bibnamefont {Oteo}},\ and\ \bibinfo
  {author} {\bibfnamefont {J.}~\bibnamefont {Ros}},\ }\bibfield  {title}
  {\bibinfo {title} {The magnus expansion and some of its applications},\
  }\href {https://doi.org/https://doi.org/10.1016/j.physrep.2008.11.001}
  {\bibfield  {journal} {\bibinfo  {journal} {Physics Reports}\ }\textbf
  {\bibinfo {volume} {470}},\ \bibinfo {pages} {151} (\bibinfo {year}
  {2009})}\BibitemShut {NoStop}%
\bibitem [{\citenamefont {Kraus}\ \emph {et~al.}(1983)\citenamefont {Kraus},
  \citenamefont {B{\"o}hm}, \citenamefont {Dollard},\ and\ \citenamefont
  {Wootters}}]{kraus-theorem}%
  \BibitemOpen
  \bibfield  {author} {\bibinfo {author} {\bibfnamefont {K.}~\bibnamefont
  {Kraus}}, \bibinfo {author} {\bibfnamefont {A.}~\bibnamefont {B{\"o}hm}},
  \bibinfo {author} {\bibfnamefont {J.~D.}\ \bibnamefont {Dollard}},\ and\
  \bibinfo {author} {\bibfnamefont {W.}~\bibnamefont {Wootters}},\ }\href@noop
  {} {\emph {\bibinfo {title} {States, effects, and operations fundamental
  notions of quantum theory: Lectures in mathematical physics at the university
  of Texas at Austin}}}\ (\bibinfo  {publisher} {Springer},\ \bibinfo {year}
  {1983})\BibitemShut {NoStop}%
\bibitem [{\citenamefont {Riesch}\ and\ \citenamefont
  {Jirauschek}(2019)}]{CPTP-long-time}%
  \BibitemOpen
  \bibfield  {author} {\bibinfo {author} {\bibfnamefont {M.}~\bibnamefont
  {Riesch}}\ and\ \bibinfo {author} {\bibfnamefont {C.}~\bibnamefont
  {Jirauschek}},\ }\bibfield  {title} {\bibinfo {title} {Analyzing the
  positivity preservation of numerical methods for the liouville-von neumann
  equation},\ }\href
  {https://doi.org/https://doi.org/10.1016/j.jcp.2019.04.006} {\bibfield
  {journal} {\bibinfo  {journal} {Journal of Computational Physics}\ }\textbf
  {\bibinfo {volume} {390}},\ \bibinfo {pages} {290} (\bibinfo {year}
  {2019})}\BibitemShut {NoStop}%
\bibitem [{\citenamefont {Drineas}\ and\ \citenamefont
  {Mahoney}(2005)}]{Nystrom-Gram}%
  \BibitemOpen
  \bibfield  {author} {\bibinfo {author} {\bibfnamefont {P.}~\bibnamefont
  {Drineas}}\ and\ \bibinfo {author} {\bibfnamefont {M.~W.}\ \bibnamefont
  {Mahoney}},\ }\bibfield  {title} {\bibinfo {title} {On the nystrom method for
  approximating a gram matrix for improved kernel-based learning},\ }\href
  {http://jmlr.org/papers/v6/drineas05a.html} {\bibfield  {journal} {\bibinfo
  {journal} {Journal of Machine Learning Research}\ }\textbf {\bibinfo {volume}
  {6}},\ \bibinfo {pages} {2153} (\bibinfo {year} {2005})}\BibitemShut
  {NoStop}%
\bibitem [{\citenamefont {Ahmed}\ \emph {et~al.}(2023)\citenamefont {Ahmed},
  \citenamefont {Quijandr\'{\i}a},\ and\ \citenamefont {Kockum}}]{GD-QPT}%
  \BibitemOpen
  \bibfield  {author} {\bibinfo {author} {\bibfnamefont {S.}~\bibnamefont
  {Ahmed}}, \bibinfo {author} {\bibfnamefont {F.}~\bibnamefont
  {Quijandr\'{\i}a}},\ and\ \bibinfo {author} {\bibfnamefont {A.~F.}\
  \bibnamefont {Kockum}},\ }\bibfield  {title} {\bibinfo {title}
  {Gradient-descent quantum process tomography by learning kraus operators},\
  }\href {https://doi.org/10.1103/PhysRevLett.130.150402} {\bibfield  {journal}
  {\bibinfo  {journal} {Phys. Rev. Lett.}\ }\textbf {\bibinfo {volume} {130}},\
  \bibinfo {pages} {150402} (\bibinfo {year} {2023})}\BibitemShut {NoStop}%
\end{thebibliography}
\end{document}